\documentclass[aip, floatfix, reprint]{revtex4-1} 

\usepackage{dcolumn}
\usepackage[utf8]{inputenc}
\usepackage[T1]{fontenc}
\usepackage{mathptmx}
\usepackage{etoolbox}
\usepackage{graphicx}
\usepackage{amsmath}
\usepackage{amssymb}
\usepackage{siunitx}
\makeatletter
\def\@email#1#2{%
 \endgroup
 \patchcmd{\titleblock@produce}
  {\frontmatter@RRAPformat}
  {\frontmatter@RRAPformat{\produce@RRAP{*#1\href{mailto:#2}{#2}}}\frontmatter@RRAPformat}
  {}{}
}%
\makeatother
\usepackage{lipsum}  

\draft 

\begin{document}

\title{Role of filler lanthanide ions on lattice dynamics of phosphide skutterudites RFe$_4$P$_{12}$ (R = La, Ce, and Pr) from first principles} 



\author{Rafaela F. S. Penacchio}
\affiliation{Institute of Physics, University of S{\~{a}}o Paulo, S{\~{a}}o Paulo, SP, Brazil}
\author{Nicholas Burns}
\affiliation{Department of Physics, University of Guelph, Guelph, ON, Canada}
\author{Maur{\'{i}}cio B. Estradiote}
\affiliation{Institute of Physics, University of S{\~{a}}o Paulo, S{\~{a}}o Paulo, SP, Brazil}
\author{Milton S. Torikachvili}
\affiliation{Department of Physics, San Diego State University, San Diego, CA, USA}
\author{Stefan W. Kycia}
\affiliation{Department of Physics, University of Guelph, Guelph, ON, Canada}
\author{S{\'{e}}rgio L. Morelh{\~{a}}o}
\email[Electronic mail: ]{morelhao@if.usp.br}
\affiliation{Institute of Physics, University of S{\~{a}}o Paulo, S{\~{a}}o Paulo, SP, Brazil}

\date{\today}

\begin{abstract}
Phosphide skutterudites primarily show promise for thermoelectric applications due to their chemical stability at high temperatures and relatively low cost. Ion doping and band gap engineering have been used to enhance their typically poor thermoelectric performance, opening avenues for practical applications. Herein, we report a comparative lattice dynamics study on the impact of filler and temperature on the structural and vibrational properties of RFe$_4$P$_{12}$ (R = La, Ce, and Pr) skutterudites. Calculations are performed within the quasi-harmonic approximation, and the results are critically compared against experimental data and other \textit{ab initio} calculations. We found gaps between the heat-carrying acoustic and optical modes, a-o gaps, of approximately 4, -2, and 0.01\,cm$^{-1}$ for La, Ce, and Pr compounds, respectively. These results suggest a filler-induced reduction in the a-o gap is attributed to the softening of the optical modes instead of the conventionally considered upward shift of acoustic modes proposed in the \textit{rattling} scenario. The distinct softening of the optical modes is rationalized by the stiffening of chemical bonds between the filler and host lattice. 
\end{abstract}


\maketitle 

Thermoelectric materials are crucial for sustainable energy solutions as they convert heat into electricity.\cite{Funahashi2021} Their efficiency is measured by the figure of merit $ZT$, inversely proportional to the lattice thermal conductivity $\kappa$. However, many high-$ZT$ materials are either too expensive or have limited stability at high temperatures.\cite{Popuri2016} In this context, filled skutterudites (SKDs) R$_{1-\delta}$T$_{4}$X$_{12}$ stand out due to their chemical stability and unusual electrical and thermal transport properties, achieving $ZT$ values as high as 1.8 at 823 K.\cite{Rogl2015} Despite having $ZT\sim 10^{-2}$ at room temperature,\cite{Watcharapasorn2000} phosphide SKDs (X = P) are particularly appealing as phosphorous is highly abundant and relatively low-cost.\cite{Quinn2023} Furthermore, doping and band gap engineering can significantly improve their thermoelectric performance, with a recent study predicting $ZT=0.82$ at 800 K for Nd-doped CeFe$_4$P$_{12}$.\cite{Limbu2022}

SKDs have a cubic structure (space group $Im\overline{3}$), with the host lattice T$_4$X$_{12}$ forming a large icosahedral cage. The progressive filling of this cage with electropositive elements R significantly reduces the lattice thermal conductivity. For example, $\kappa$ drops over one order of magnitude when filling the Co$_4$Sb$_{12}$ structure with Nd and by a factor of 6-7 when the complete filling is achieved in the Ce$_{1-\delta}$Fe$_4$Sb$_{12}$ structure.\cite{Kuznetsov2003,Koza2010} This suppression of $\kappa$ was initially rationalized as the filler ion R \textit{rattling} independently in the cage, acting as individual, non-dispersive, low-energy oscillators and giving rise to additional scattering channels for heat-carrying acoustic phonons.\cite{Keppens1998} Presently, it is well understood that the vibrational properties of most filled SKDs subsystems arise from correlated filler-cage dynamics, particularly true for antimonide SKDs (T = Sb).\cite{Koza2008, Koza2015, Abrantes2022}

Although phosphide SKDs RFe$_4$P$_{12}$ have been extensively studied due to interesting phenomena such as superconductivity, antiferromagnetism, valence fluctuation, and filler atom dependent metal-insulator transitions,\cite{Torikachvili1987} the role of filler ions in the vibrational dynamics remains controversial. Experimental evidence supporting the \textit{rattling} scenario is largely based upon large atomic displacements (B-factors) obtained from diffraction data,\cite{Grandjean1984, Jeitschko1977, Keller2001} Einstein temperatures obtained from EXASF studies,\cite{Cao2004} and frequencies from infrared and optical reflectance spectroscopy.\cite{Dordevic1999} Conversely, other results support a more complex filler-cage interaction such as the absence of second-order phonons in Raman spectroscopy measurements,\cite{Kojima2007, Ogita2007} no site-dependence with the spin-lattice relaxation rate in La-NMR and P-NMR/Sb-NQR measurements,\cite{Nakai2007} and relative atomic displacements from multiwavelength dynamical diffraction data.\cite{Valerio2020} On top of this, most current first-principles modeling on these materials has been performed at fixed temperatures, mainly focusing on the 0\,K equilibrium structure. Such modeling leaves room for more comprehensive and systematic computational studies of the effect of temperature on the properties of these materials.\cite{Hasegawa2008,Xu2007,Hachemaoui2010,Ameri2014, Abdelakader2024} In this work, a comparative lattice-dynamics study is carried out on the impact of both filler atom and temperature on the structural and vibrational properties of RFe$_4$P$_{12}$ (R = La, Ce, and Pr) SKDs. Among the main results are the distinct gaps between the acoustic- and optical-phonon modes, the a-o gap, depending on the filler,  as well as the thermal expansion coefficients of the compounds, which were experimentally verified in powder samples from 80\,K  to room temperature. 

Density functional theory (DFT) calculations were performed using the Quantum Espresso distribution\cite{Giannozzi2009, Giannozzi2017} at the generalized gradient approximation (GGA) level. Modeling structural, vibrational, and electronic properties were based on the Perdew-Burke-Ernzerhof (PBE) exchange-correlation functional,\cite{Perdew1996} together with the standard solid-state pseudopotentials (SSSP) optimized for precision\cite{Prandini2018}. A kinetic energy cutoff of 1360\,eV was sufficient in producing convergence in the total energy up to 2.5 meV/atom for a given \textit{k}-point sampling. A $6\times6\times6$ $\Gamma$-centered Monkhorst-Pack mesh was applied for structural calculations, in addition to a Gaussian smearing width of 0.1\,eV for La- and Pr-filled SKDs. 

Lattice-dynamics calculations were carried out using Phonopy within the finite-displacement supercell approach.\cite{Togo2023} A $2\times2\times2$ supercell (containing 136 atoms) sampling only the $\Gamma$-point was employed. During post-processing, thermodynamic properties were calculated by sampling a $100\times100\times100$ $\Gamma$-centered $q$ mesh. The temperature dependence of lattice constants, bulk moduli, and specific heat were calculated by the quasi-harmonic approximation (QHA) as implemented in Phonopy. For the QHA calculations, additional finite-displacement calculations were performed on unit cells at approximately $\pm$2.5\% of the equilibrium volume in steps of 0.5\%. After computing the thermal expansion, further calculations were performed on structures with volumes corresponding to 4, 80, 180, 320, and 500\,K.

X-ray scattering experiments were carried out at the Brockhouse Diffraction Sector beamlines of the Canadian Light Source.\cite{Gomez2018} Resonant diffraction in single-crystals, near the Fe absorption edge,\cite{Penacchio2023} at the wiggler low energy (WLE) beamline;\cite{Leontowich2021} and total scattering as a function of temperature in powder samples, with 55\,keV X-rays, at the wiggler high energy (WHE) beamline. Specifics about the total scattering data acquisition and processing are detailed in recent work.\cite{Burns2023} All RFe$_4$P$_{12}$ samples were originally obtained as high-quality single-crystals synthesized by the tin-flux method.\cite{Torikachvili1987} Temperature values were calibrated through a NIST nickel powder sample, while the lattice constants were determined by Rietveld analysis using the GSAS-II software package.\cite{Toby2013} In single-crystals, the lattice constants were obtained by the azimuthal scan method.\cite{la04,rf07,am10,Valerio2020} 

\begin{figure}
\includegraphics[width=.4\textwidth]{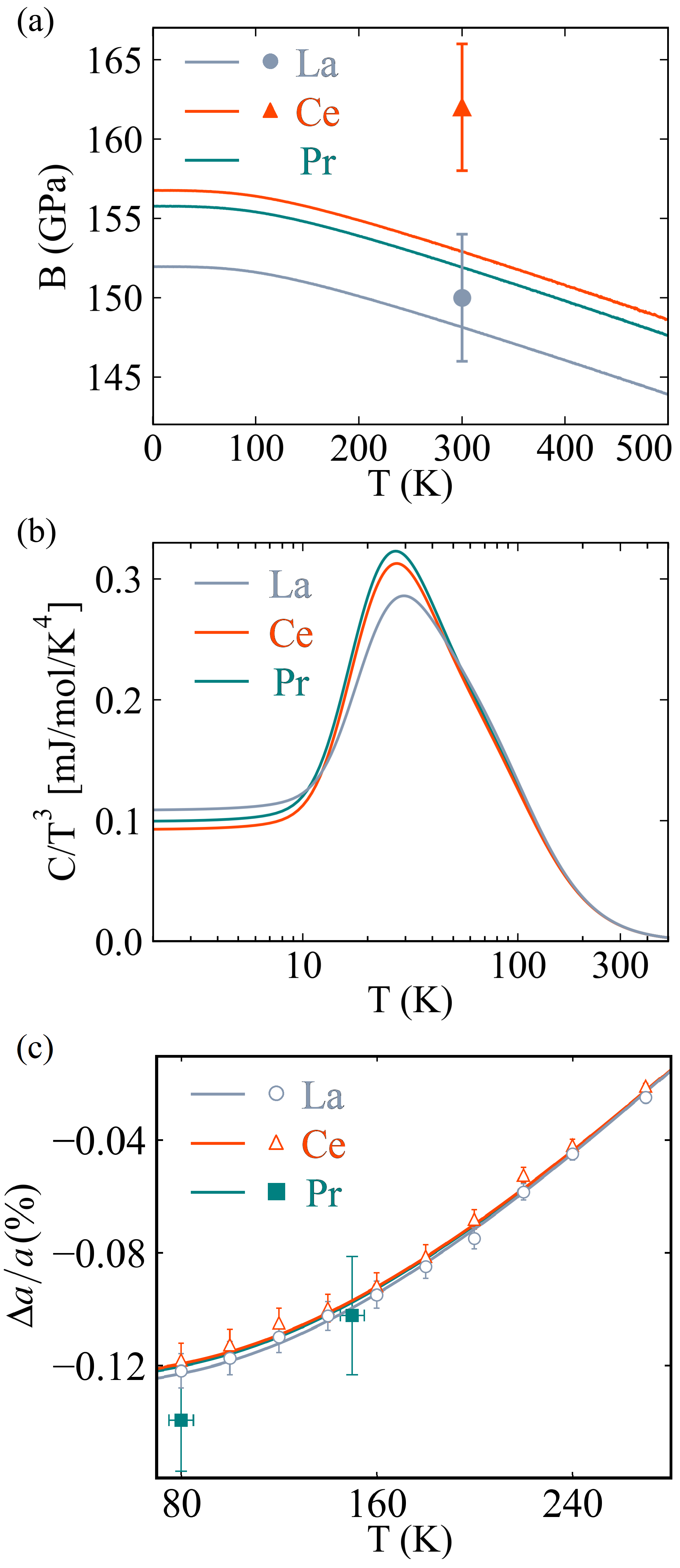}
\caption{\label{fig:DFTresults} DFT results (solid lines) as a function of temperature $T$ in LaFe$_{4}$P$_{12}$ (La), CeFe$_{4}$P$_{12}$ (Ce), and PrFe$_{4}$P$_{12}$ (Pr) SKDs. (a) Bulk modulus $B$. Additionally, few experimental values (symbols) available in the literature are also shown.\cite{Shirotani2004, Hayashi2010}. (b) Lattice specific heat $C$, given in terms of the ratio $C/T^3$. (c) Relative lattice constant variation $\Delta a/a$ together with experimental values from X-ray scattering in powder (open symbols) and single-crystal (closed symbols) samples.}
\end{figure}

\begin{table*}[htbp]
\caption{Lattice constants $a$, Einstein $\theta_E$ and Debye $\theta_D$ temperatures, as obtained from DFT calculations. Experimental values of $a$ are from X-ray scattering in powder and single-crystal (s.-c.) samples, while $\theta_E$ and $\theta_D$ are from literature. Uncertainties are presented in parentheses.}\label{tab:StrucProp}
\begin{ruledtabular}
    \begin{tabular}{ccccccccc}
         & \multicolumn{4}{c}{DFT} & \multicolumn{4}{c}{Experimental}  \\
         \cline{2-5}
         \cline{6-9}  
         & \multicolumn{2}{c}{$a$ (\r{A})} & & & \multicolumn{2}{c}{$a$ (\r{A}) ($\sim$ 300\,{\rm K})}  &  & \\
         \cline{2-3} \cline{6-7}
   SKD      & 0\,K & 296\,K & $\theta_E ({\rm K})$ & $\theta_D ({\rm K})$ & powder\;\cite{Jeitschko1977} & s.-c. & $\theta_E ({\rm K})$ & $\theta_D ({\rm K})$ \\ \hline
LaFe$_4$P$_{12}$ & 7.8489 & 7.8587 & 145 & 662 & 7.8316(5) & 7.8338(14) & 147(1)\,\cite{Matsuhira2009,Mizumaki2011} & 670\;\cite{Matsuhira2009} \\
CeFe$_4$P$_{12}$ & 7.8095 & 7.8189 & 135 & 694 & 7.7920(10) & 7.7914(3) & 148(5)\,\cite{Cao2004}/142(1)\,\cite{Mizumaki2011}& ---\\
PrFe$_4$P$_{12}$ & 7.8053 & 7.8149 & 134 & 678 & 7.8149(9) & 7.8116(7) & 137(1)\,\cite{Mizumaki2011} & ---\\
    \end{tabular}
\end{ruledtabular}
\end{table*}

Main results from \textit{ab initio} calculations as a function of temperature $T$ and filler R are presented in Fig.~\ref{fig:DFTresults}. The bulk moduli $B$, Fig. \ref{fig:DFTresults}(a), decreases by about 5\% between 0 and 500\,K for all three SKDs. At 300\,K, the calculated $B$ values for La-, Ce- and Pr-filled SKDs are 148, 153, and 152\,GPa, respectively. The former is in good agreement with the experimentally measured value of 150\,GPa,\cite{Hayashi2010} while the value obtained for the Ce-SKD deviates by about 5\% from experiment.\cite{Shirotani2004} For the Pr compound, the experimental bulk modulus of 211\,GPa at 77\,K, not shown in Fig. \ref{fig:DFTresults}(a), exceeds the calculated value by approximately 35\%.\cite{Nakanishi2001} Comparatively, calculations within local density approximation (LDA) deviate from experimental values by 3 to 18\%. 

Calculated values of lattice specific heat $C$ are presented in Fig. \ref{fig:DFTresults}(b). Absence of the superconducting transition in La-SKD at $4.7$\,K and magnetic ordering in Pr-SKD below 6.2\,K is a result of the limited capability of standard DFT calculations at the GGA level in describing the strong electron correlations involved in these transitions.\cite{Cohen2012, Pavarini2021} Peak positions in Fig. \ref{fig:DFTresults}(b) were used to estimate the Einstein temperatures $\theta_E$ of La-, Ce-, and Pr-filled SKDs, resulting in 145, 135, and 134\,K, respectively. These values are all within 10\% of experimental data,\cite{Matsuhira2009, Cao2004, Mizumaki2011} as shown in Table~\ref{tab:StrucProp}. The $C(T)$ curve below 7.5\,K is well-fit by the Debye $T^3$ law, yielding Debye temperatures $\theta_D$ of 662, 694, and 678\,K, respectively. For the La-SKD, the obtained result agrees well with the experimentally measured value of 670\,K available in the literature;\cite{Matsuhira2009} there is a lack of experimental data for the other two SKDs. It should be noted that the Debye temperatures are behaving differently than the expected trend $\theta_D\propto M^{-1/3}$ with molar mass $M$.\cite{Poirier1989} The La-, Pr-, and Ce-SKDs exhibit $\theta_D$ values in ascending order, suggesting a corresponding increase in chemical bond strength and sound velocities.

Theoretical and experimental relative variations of lattice constants as a function of $T$ are shown in Fig. \ref{fig:DFTresults}(c). Table~\ref{tab:StrucProp} reports the absolute values of the lattice constants at 0\,K and 296\,K (room temperature), alongside with additional theoretical and experimental values. At 296\,K, the lattice constants of La- and Ce-SKDs are 0.35\% larger than the experimental values,\cite{Jeitschko1977} as expected for GGA functionals.\cite{Haas2009} Conversely, the lattice constant of Pr-SKD is in perfect agreement with the literature values. The calculated values, in Table~\ref{tab:StrucProp} (column 3), are much closer to the experimentally measured values than other previous calculations at LDA and LSDA (local spin density approximation) levels.\cite{Benalia2008, Hachemaoui2010} Both LDA and LSDA underestimated 0\,K lattice constants by more than 1\%. The QHA calculations effectively reproduce the thermal expansion of La- and Ce-SKDs in the temperature range of 80 to 300 K, as seen in Fig. \ref{fig:DFTresults}(c). Lattice constants from single crystal samples at room temperature are also presented in Table~\ref{tab:StrucProp} (column 7) for the sake of comparison. Despite the $\Delta a/a (T)$ curves for Ce- and Pr-SKDs being nearly coincident, more accurate experimental data points are needed to confirm the latter. Nonetheless, at room temperature, the QHA calculations reproduce the absolute values of the lattice constants of these filled SKDs reasonably well. 

\begin{figure*}[ht]
\includegraphics[width=\textwidth]{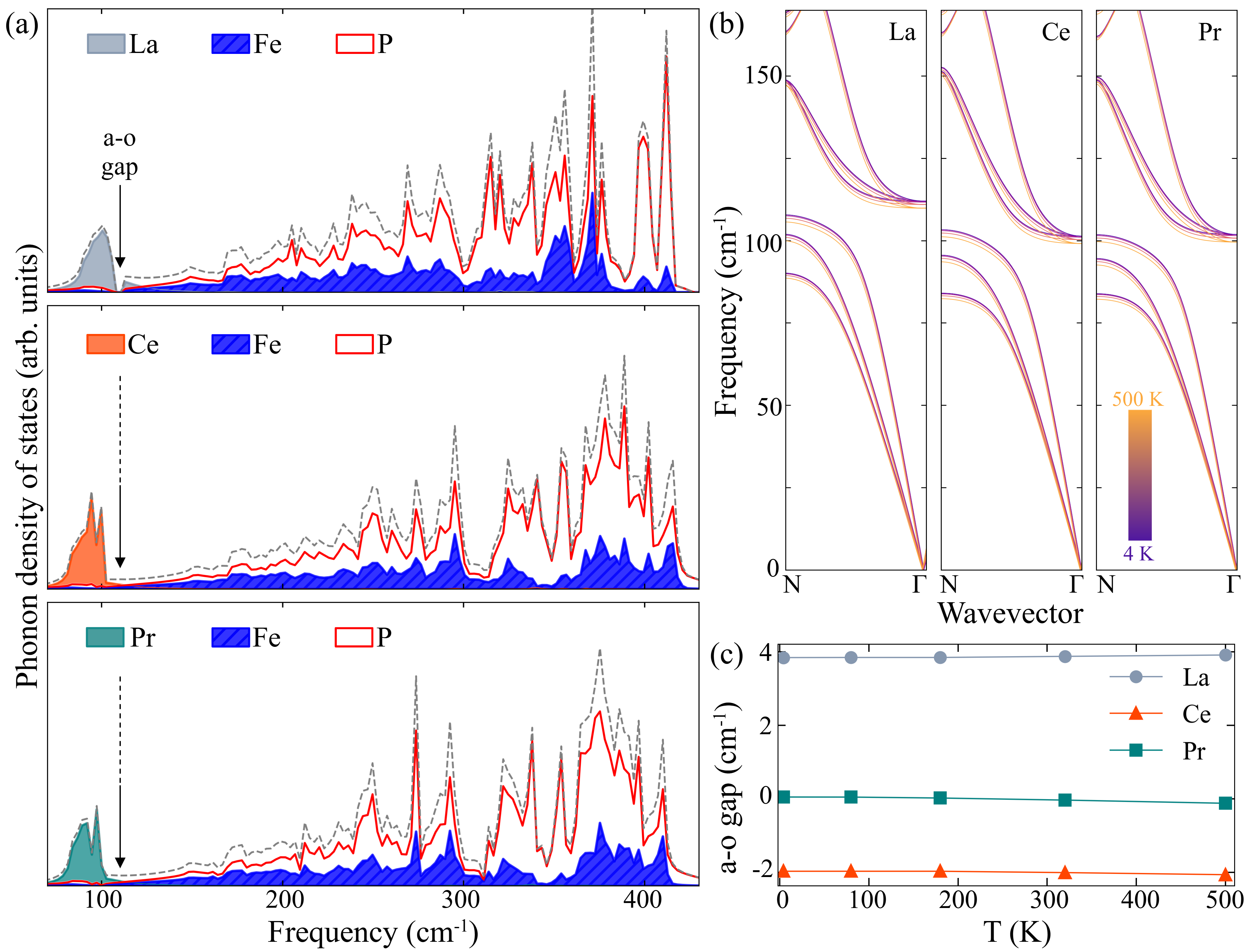}
\caption{\label{fig:PhononBands} Vibrational properties obtained from \textit{ab initio} lattice-dynamics calculations in La-, Ce-, and Pr-filled SKDs. (a) Partial (colored lines with shaded areas) and total (dashed lines) phonon DOS at T = 320\,K, as indicated on each panel. The a-o gap (arrows) at around 110\,cm$^{-1}$ is also indicated. (b) Temperature dependence of the low-energy phonon branches around the a-o gap in the range 4\,K (purple) to 500\,K (orange). (c) A-o gap as a function of temperature.}
\end{figure*}

The phonon density of states (DOS) at 320\,K in Fig.~\ref{fig:PhononBands}(a) is representative, within QHA, of the evaluated temperature range. Increasing temperature only results in a downward shift of both acoustic and optical modes for all three SKDs, as shown in Fig.~\ref{fig:PhononBands}(b); for additional information please refer to the Supplementary Material (SM). In general, vibrational modes associated with the fillers are located in the low-frequency range below 100\,cm$^{-1}$, contributing predominantly to three acoustic branches, consistent with the scenario reported for heavy-filler antimonide skutterudites.\cite{Koza2010} Regardless of the filler R, the DOS for P and Fe atoms are widely distributed in the range of phonon frequencies above 100\,cm$^{-1}$, showing an overlap across the entire spectrum due to the strong covalent bond between these atoms. However, in La-SKD, P and Fe vibrations are almost totally decoupled from each other at high frequencies, above 380\,cm$^{-1}$ as presented in Fig.~\ref{fig:PhononBands}(a) (top panel). This result can be correlated to the smaller Debye temperature obtained for this compound, Table~\ref{tab:StrucProp} (column 5).

Although the filler R predominantly contributes to the acoustic modes, the vibrational dynamics of the host lattice Fe$_4$P$_{12}$ depend upon the filler type, as demonstrated in Fig. \ref{fig:PhononBands}(a). For instance, the DOS at high frequencies is associated with cage vibrational modes yet still undergoes significant changes as a function of the filler, mainly for La with respect to Ce and Pr. Another distinct filler-dependent feature is the softening of optical modes that directly affects the gap between acoustic and optical modes, as highlighted in Fig. \ref{fig:PhononBands}(b) along the $\Gamma-N$ path. This softening is particularly pronounced in the Ce-SKD, leading to a negative indirect a-o gap, where the lower energy optical mode is less energetic than the higher energy acoustic mode. In contrast, the weaker softening in La-SKD implies a positive gap, while the Pr-SKD falls between the two, exhibiting a null gap. As a function of temperature, the a-o gaps are nearly constant, which is observed in Fig. \ref{fig:PhononBands}(c). Null or negative gaps favor phonon scattering from heat-carrying acoustic to optical branches, leading to a reduction in lattice thermal conductivity. This result suggests that the Ce-filled SKD is more promising for thermoelectric applications than isotypic La- or Pr-SKDs, as previously highlighted when considering the transport properties of these materials \cite{Sato2000, Dordevic1999, Watcharapasorn2000}. For example, the Ce-SKD is a semiconductor and its Seebeck coefficient has been observed to increase with temperature, almost linearly from 58\,$\mu$V/K at room temperature to 128\,$\mu$V/K at near 900\,K.\cite{aw00} 

Distinct a-o gaps, as shown in Fig. \ref{fig:PhononBands}(b,c), are caused by changes in force constants of the filler-cage. La exhibits only the 3+ valency without hybridization of $p-f$ orbitals, while Ce can manifest a mix of 3+ and 4+ valences, each associated with different chemical bonding.\cite{Grosvenor2006, Rayjada2010} Similarly, Pr also presents mixed 3+ and 4+ valences with $p-f$ hybridization in both instances.\cite{Yamasaki2005, Matsunami2008} Further details on the electronic density of states are provided in SM. It is worth mentioning that the hybridization of $4f$-orbitals with the conduction band is also suggested to be responsible for the unusual temperature dependence of the transport properties of the Ce-SKD.\cite{Sato2000}

\begin{figure}
\includegraphics[width=.5\textwidth]{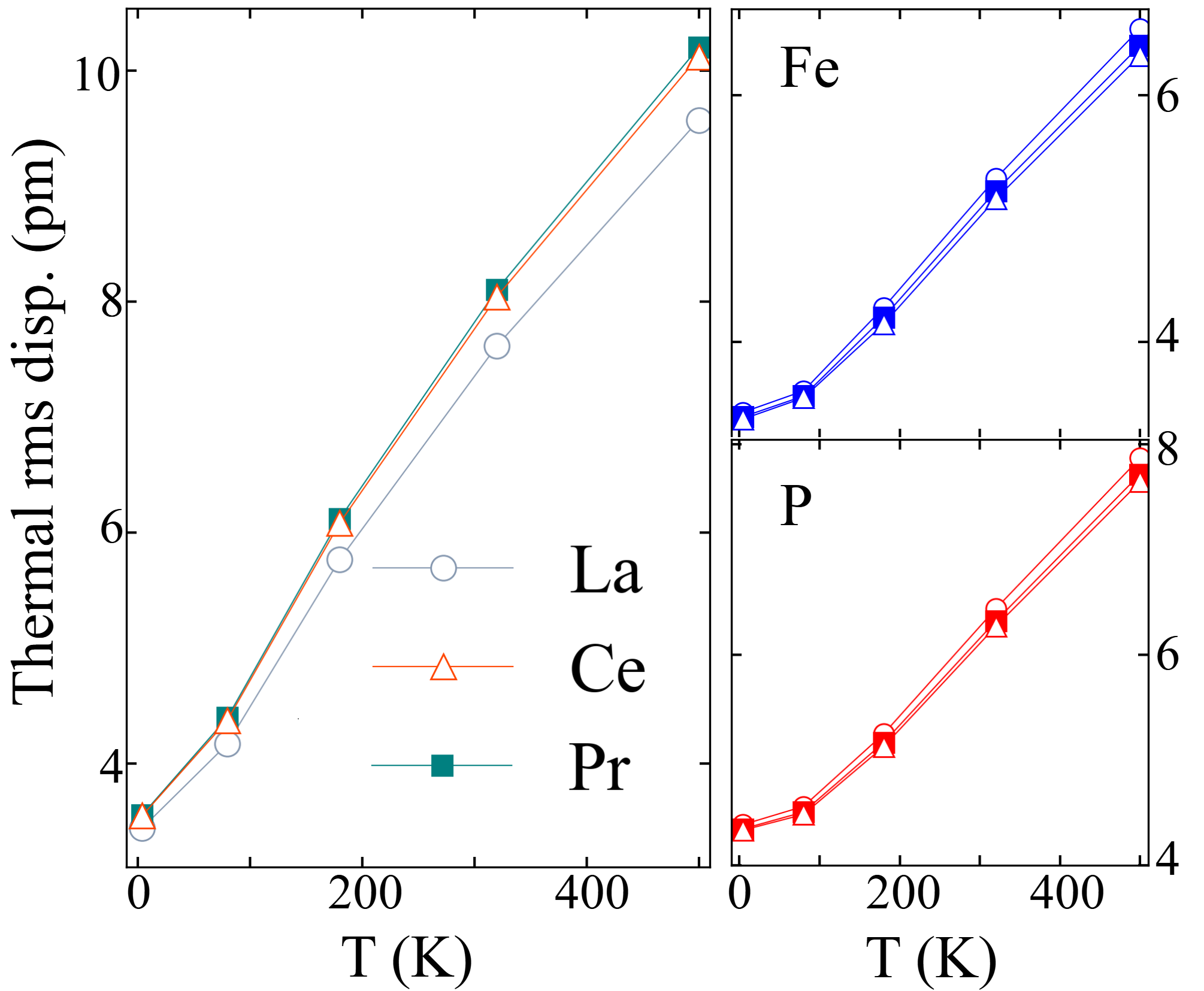}
\caption{\label{fig:RMSD} Atomic root-mean-square displacements (rms disp.) as a function of temperature in filled SKDS: LaFe$_4$P$_{12}$ (circles), CeFe$_4$P$_{12}$ (triangles), and PrFe$_4$P$_{12}$ (squares). Isotropic values for fillers (left panel) and host lattice atoms, Fe (right top panel) and P (right bottom panel).}
\end{figure}

Lastly, Fig. \ref{fig:RMSD} shows the calculated isotropic root-mean-square (rms) displacement as a function of temperature. The values obtained for the filler R ($=$ La, Ce, and Pr) at room temperature agree well with reported experimental data from Rietveld refinement within a tolerance of 7\%.\cite{Jeitschko1977, Grandjean1984, Keller2001} This close correspondence was also verified for the rms displacements of Fe and P atoms in La- and Ce-filled SKDs. Conversely, in the Pr-SKD, the calculated values for Fe and P atoms exceed the experimental data by up to 28\%. According to the calculations presented here, Fe atoms always have the smaller rms displacement regardless of filler and temperature. Moreover, cage P atoms have thermal motion larger than the fillers at low temperatures below 140\,K for the La-SKD, and 86\,K for both Ce- and Pr-SKDs. In other words, QHA calculations predict that, at room temperature, the thermal motion of the filler ions has a larger amplitude than that of the cage atoms, in agreement with experimental data.

In summary, our comparative lattice-dynamics study of RFe$_4$P$_{12}$ (R = La, Ce, and Pr) SKDs has effectively reproduced the thermal expansion of La and Ce compounds. Also, it has predicted values for the bulk modulus at room temperature, as well as the Debye and Einstein temperatures, with reasonable accuracy. By considering the phonon frequencies and thermal displacements within the quasi-harmonic approximation, we find out that the vibrational dynamics of the host lattice are mainly affected by the chemical properties of the filler ions. The filler-induced stiffening of chemical bonds, as evidenced by an increase in Debye temperature, results in a downshift shift of the energy of the optical modes with a consequent reduction in the a-o gap, which contains the heat-carrying acoustic phonons. Such a trend is the opposite situation to the expected \textit{rattling} scenario, where phonon scattering is favored by an increase in the energy of acoustic modes rather than the softening of optical modes. Furthermore, the a-o gaps are considered to be nearly constant as a function of temperature for all three compounds evaluated here, suggesting that the reduction in lattice thermal conductivity through phonon scattering is more pronounced in the Ce-filled SKD, which exhibits the smallest gap. In terms of material design, this work provides fruitful insights into strategies for improving the thermoelectric performance of phosphide skutterudites. 

\section*{Supplementary Material}

See the supplementary material for the temperature dependence of phonon dispersion curves, electronic density of states at 320\;K, experimental results from X-ray total scattering in powder samples, and dynamical diffraction in single crystals.

This work was supported by the Fundação de Amparo à Pesquisa do Estado de São Paulo - FAPESP and by the National Council for Scientific and Technological Development - CNPq.


\begin{acknowledgments}
We acknowledge the Canadian Light Source, a National Research Facility of the University of Saskatchewan, which is supported by the Canada Foundation for Innovation (CFI), the Natural Sciences and Engineering Research Council (NSERC), the National Research Council (NRC), the Canadian Institutes of Health Research (CIHR) and the Government of Saskatchewan. 

\end{acknowledgments}

\section*{Conflict of Interest}

The authors have no conflicts to disclose.


\section*{Data Availability}

The data that support the findings of this study are available from the corresponding author upon reasonable request.

\bibliography{bibliography}

\end{document}